\documentclass[10pt]{article}
\usepackage{graphicx}
\usepackage{amsmath}
\usepackage{amssymb}
\usepackage{caption2}
\begin{document}
\bibliographystyle{prsty}
\begin{center}
{\large {\bf \sc{  The semileptonic decays  $B_c^* \to \eta_c \ell \bar{\nu}_{\ell} $ with QCD sum rules }}} \\[2mm]
Zhi-Gang Wang \footnote{E-mail,zgwang@aliyun.com.  }     \\
 Department of Physics, North China Electric Power University,
Baoding 071003, P. R. China
\end{center}

\begin{abstract}
In this article, we calculate the $B_c^* \to \eta_c$ form-factors with the three-point QCD sum rules, then study   the semileptonic decays  $B_c^* \to \eta_c \ell \bar{\nu}_{\ell}$. The tiny decay widths may be observed   experimentally  in the future at the LHCb, while
 the $B_c^* \to \eta_c$ form-factors can be taken as basic input parameters in other  phenomenological analysis.
\end{abstract}

 PACS number: 12.38.Lg, 13.20.Jf

Key words: $B_c^*$-meson decays, QCD sum rules, Semileptonic decays

\section{Introduction}

The bottom-charm  quarkonium states  are of special interesting, the ground states $B_c$ and $B_c^*$ which lie below the $BD$, $BD^*$, $B^*D$, $B^*D^*$ thresholds cannot annihilate into gluons, and decay weakly through  $\bar{b} \to \bar{c}W^+$, $c \to s W^+$, $c\bar{b}\to W^+$ at the quark level, furthermore, the $B_c^*$ mesons  also have the radiative transitions $B_c^* \to B_c \gamma$. The $B_c^{\pm}$ mesons have measurable lifetime, while the $B_c^{*\pm}$ mesons would have widths less than a hundred $\rm{KeV}$ \cite{GI}.  The semileptonic decays $B_c^{\pm} \to J/\psi \ell^{\pm}\bar{\nu}_{\ell}$, $B_c^{+} \to J/\psi e^{+}\bar{\nu}_{e} $ were   used to measure the $B_c$ lifetime and the hadronic decays $B_c^{\pm} \to J/\psi \pi^{\pm}$ were used to measure the $B_c$ mass in $p\bar{p}$ collisions at the energy  $\sqrt{s}=1.96\,\rm{TeV}$ by the CDF and D0 collaborations \cite{CDF-life,D0-life,CDF2008,D02008}. Now the average values are $\tau_{B_c}=(0.45 \pm 0.04)\times 10^{-12}\,s$ and  $m_{B_c}=(6.277 \pm 0.006)\,\rm{GeV}$  from the Particle Data Group \cite{PDG}. The $B_c^*$ mesons have not been observed yet, but they are expected to be observed  and their properties be studies in details at the large hadron collider (LHC).   The LHC will be the world's most
copious  source of the $b$ hadrons, and  a complete spectrum of the
$b$ hadrons will be available through gluon fusion. In proton-proton
collisions at $\sqrt{s}=14\,\rm{TeV}$, the $b\bar{b}$ cross section is expected to be $\sim 500\mu b$ producing $10^{12}$
$b\bar{b}$ pairs in a standard  year of running at the LHCb
operational luminosity of $2\times10^{32} \rm{cm}^{-2}
\rm{sec}^{-1}$ \cite{LHC}.

The semileptonic decays $b \to c \ell \bar{\nu}_{\ell} $ are excellent subjects
in exploring the CKM matrix element $V_{cb}$, we can use both the exclusive and inclusive $b \to c$ transitions to study the CKM matrix element $V_{cb}$.  The semileptonic and nonleptonic $B_c$-decays have been studied extensively \cite{HQP2004}, in those studies, we often encounter the
 $B_c \to P,\,V$ form-factors, which are highly nonperturbative quantities and should be calculated by some nonperturbative theoretical approaches.
 In this article, we calculate the $B_c^* \to \eta_c$ form-factors with the three-point QCD sum rules, then take those form-factors as basic  input parameters
 to study the semileptonic decays $B_c^*\to \eta_c \ell \bar{\nu}_{\ell}$. The  QCD sum rules is a powerful nonperturbative theoretical tool in studying the
ground state hadrons, and has given a lot of successful descriptions of the hadron properties  \cite{SVZ79,Reinders85,QCDSR-review,NarisonBook}.   There have been several works on the
semileptonic   $B_c$-decays with the three-point  QCD sum rules  \cite{Colangelo-BC,Narison-BC,Coulomb-BC,QCDSR-BC}, while there does not exist work on the  semileptonic $B^*_c$-decays.

The article is arranged as follows:  we study the $B_c^* \to \eta_c  $ form-factors  using
  the three-point QCD sum rules in Sect.2; in Sect.3, we present the numerical results and discussions; and Sect.4 is reserved for our
conclusions.

\section{ The $B_c^* \to \eta_c$ form-factors with QCD sum rules}
We study the $B_c^* \to \eta_c$ form-factors with  the three-point correlation function $\Pi_{\mu\nu}(p_1,p_2)$,
\begin{eqnarray}
\Pi_{\mu\nu}(p_1,p_2)&=&i^2\int d^4 x d^4 y e^{ip_2\cdot x-ip_1\cdot y} \langle 0|T\{J_5(x) j_\mu (0) J_\nu(y)\} |0\rangle \, ,
\end{eqnarray}
where
\begin{eqnarray}
J_5(x)&=&\bar{c}(x)i\gamma_5c(x)\, , \nonumber\\
j_\mu(0)&=&\bar{c}(0)\gamma_\mu(1-\gamma_5)b(0)\, , \nonumber\\
J_\nu(y)&=&\bar{b}(y)\gamma_\nu c(y)\, ,
\end{eqnarray}
the pseudoscalar current $J_5(x)$ and vector current $J_\nu(y)$  interpolate the pseudoscalar meson $\eta_c$ and vector meson $B_c^*$, respectively,
the $j_\mu(0)$ is the transition chiral current sandwiched between the $B_c^*$ and $\eta_c$ mesons.

We can insert  a complete set of intermediate hadronic states with
the same quantum numbers as the current operators $J_5(x)$ and $J_\nu(y)$ into the
correlation function $\Pi_{\mu\nu}(p_1,p_2)$  to obtain the hadronic representation
\cite{SVZ79,Reinders85}. After isolating the ground state
contributions come from the heavy mesons $B_c^*$ and $\eta_c$ , we get the following result,
\begin{eqnarray}
\Pi_{\mu\nu}(p_1,p_2)&=&\frac{\langle 0|J_5(0) |\eta_c(p_2)\rangle \langle\eta_c(p_2)|j_\mu(0) |B_c^*(p_1)\rangle \langle B_c^*(p_1)  |J_\nu(0) |0   \rangle}{(m_{B_c^*}^2-p_1^2)(m_{\eta_c}^2-p_2^2)} +\cdots   \, ,\nonumber\\
&=&\frac{f_{\eta_c}m_{\eta_c}^2f_{B_c^*}m_{B_c^*}}{2m_c(m_{B_c^*}^2-p_1^2)(m_{\eta_c}^2-p_2^2)}\left\{ -ig_{\mu\nu}(m_{B_c^*}+m_{\eta_c})A_1(q^2)
+ip_{1\mu}p_{2\nu}\frac{A_{+}(q^2)+A_{-}(q^2)}{m_{B_c^*}+m_{\eta_c}}\right. \nonumber\\
&&\left.+ip_{2\mu}p_{2\nu}\frac{A_{+}(q^2)-A_{-}(q^2)}{m_{B_c^*}+m_{\eta_c}} -\epsilon_{\mu\nu\alpha\beta}p_1^\alpha p_2^\beta \frac{2V(q^2)}{m_{B_c^*}+m_{\eta_c}}+\cdots\right\}+\cdots \, ,
\end{eqnarray}
where we have used the following definitions for the $B_c^* \to \eta_c$ form-factors and decay constants of the $B_c^*$ and $\eta_c$ mesons,
\begin{eqnarray}
\langle\eta_c(p_2)|j_\mu(0) |B_c^*(p_1)\rangle&=& i\varepsilon_\mu (m_{B_c^*}+m_{\eta_c})A_1(q^2)+i(p_1+p_2)_\mu \varepsilon \cdot q \frac{A_{+}(q^2)}{m_{B_c^*}+m_{\eta_c}} \nonumber\\
&&+iq_\mu \varepsilon \cdot q \frac{A_{-}(q^2)}{m_{B_c^*}+m_{\eta_c}}+\epsilon_{\mu\nu\alpha\beta}\varepsilon^{\nu}p_1^\alpha p_2^\beta \frac{2V(q^2)}{m_{B_c^*}+m_{\eta_c}}\, ,\\
\langle0|J^{\dagger}_\mu(0) |B_c^*(p_1)\rangle&=&f_{B_c^*}m_{B_c^*}\varepsilon_\mu \, ,\nonumber\\
\langle0|J_5(0) |\eta_c(p_2)\rangle&=&\frac{f_{\eta_c}m_{\eta_c}^2}{2m_c} \, ,
\end{eqnarray}
$q_\mu=(p_1-p_2)_\mu$, the $\varepsilon_\mu$ is the polarization  vector of the $B_c^*$ meson and satisfies the relation,
\begin{eqnarray}
\sum_{\lambda}\varepsilon_{\mu}^*(\lambda,p)\varepsilon_{\nu}(\lambda,p)&=&-g_{\mu\nu}+\frac{p_\mu p_\nu}{p^2} \, .
\end{eqnarray}
In this article, we choose the tensor structures $g_{\mu\nu}$, $p_{1\mu}p_{2\nu}$,  $p_{2\mu}p_{2\nu}$ and $\epsilon_{\mu\nu\alpha\beta}p_1^{\alpha}p_2^{\beta}$ to
study the weak form-factors.

Here we will take a short digression to discuss the relations among the form-factors based on the heavy quark symmetry \cite{HQS}.  The $B_c^* \to \eta_c$ form-factors can be rewritten as
\begin{eqnarray}
\langle\eta_c(p_2)|j_\mu(0) |B_c^*(p_1)\rangle&=& i\varepsilon_\mu (m_{B_c^*}+m_{\eta_c})A_1(q^2)+i(p_1+p_2)_\mu \varepsilon \cdot q \frac{A_{2}(q^2)}{m_{B_c^*}+m_{\eta_c}} \nonumber\\
&&-2m_{B_c^*}iq_\mu \frac{\varepsilon \cdot q}{q^2} \left[A_{3}(q^2)-A_{0}(q^2) \right]+\epsilon_{\mu\nu\alpha\beta}\varepsilon^{\nu}p_1^\alpha p_2^\beta \frac{2V(q^2)}{m_{B_c^*}+m_{\eta_c}}\, ,
\end{eqnarray}
where
\begin{eqnarray}
A_3(q^2)&=&\frac{m_{B_c^*}+m_{\eta_c}}{2m_{B_c^*}}A_1(q^2)+\frac{m_{B_c^*}-m_{\eta_c}}{2m_{B_c^*}}A_2(q^2)\, ,\nonumber\\
A_{+}(q^2)&=&A_2(q^2)\, , \,\, A_3(0)=A_0(0)\, ,\nonumber\\
A_{-}(q^2)&=&-2m_{B_c^*}(m_{B_c^*}+m_{\eta_c})\frac{A_{3}(q^2)-A_{0}(q^2)}{q^2} \, .
\end{eqnarray}
In the heavy quark limit, the $B_c^* \to \eta_c$ form-factors can be parameterized by the universal Isgur-wise function $\xi(\omega)$,
\begin{eqnarray}
\langle\eta_c(v^{\prime})|j_\mu(0) |B_c^*(v)\rangle&=& i\left[\varepsilon_\mu (v\cdot v^{\prime}+1)-v_\mu \varepsilon \cdot v^{\prime}\right]\xi(\omega)+\epsilon_{\mu\nu\alpha\beta}\varepsilon^{\nu} v^\alpha v^{\prime\beta} \xi(\omega)\, ,
\end{eqnarray}
where the $v_\mu$ and $v_\mu^{\prime}$ are four-velocities, and $\omega=v \cdot v^{\prime}$. Then we obtain the following relations,
\begin{eqnarray}
V(q^2)&=&A_2(q^2)=A_0(q^2)=A_1(q^2)\left[1-\frac{q^2}{(m_{B_c^*}+m_{\eta_c})^2}\right]^{-1}=\frac{m_{B_c^*}+m_{\eta_c}}{2\sqrt{m_{B_c^*} m_{\eta_c}}}\xi(\omega)\, .
\end{eqnarray}
The vector state $|B_c^*(v)\rangle$ relates with the pseudoscalar state $|B_c(v)\rangle$ through
$|B_c^*(v)\rangle=2S_b^3|B_c(v)\rangle$, where the $S_b^3$ is the heavy quark spin operator.  We can also express the $B_c \to \eta_c$ form-factors in terms of the
 Isgur-wise function $\xi(\omega)$,
\begin{eqnarray}
\langle\eta_c(v^{\prime})|j_\mu(0) |B_c(v)\rangle&=&  \xi(\omega)(v+v^{\prime})_{\mu}\, .
\end{eqnarray}
On the other hand, the $B_c \to \eta_c$ form-factors are usually parameterized by the two form-factors $F_1(q^2)$ and $F_0(q^2)$,
\begin{eqnarray}
\langle\eta_c(p_2)|j_\mu(0) |B_c(p_1)\rangle&=&  F_{1}(q^2)\left[(p_1+p_2)_{\mu}-\frac{m^2_{B_c}-m_{\eta_c}^2}{q^2}q_\mu\right]+F_{0}(q^2)\frac{m^2_{B_c}-m_{\eta_c}^2}{q^2}q_\mu\, .
\end{eqnarray}
The  form-factors $F_1(q^2)$ and $F_0(q^2)$ relate with the Isgur-wise function $\xi(\omega)$ through,
\begin{eqnarray}
F_1(q^2)&=&F_0(q^2)\left[1-\frac{q^2}{(m_{B_c}+m_{\eta_c})^2}\right]^{-1}=\frac{m_{B_c}+m_{\eta_c}}{2\sqrt{m_{B_c} m_{\eta_c}}}\xi(\omega)\, .
\end{eqnarray}
Finally we obtain the following relations among the $B_c^*\to \eta_c$ and $B_c \to \eta_c$ form-factors in the heavy quark limit,
\begin{eqnarray}
V(q^2)&=&A_2(q^2)=A_0(q^2)=F_1(q^2)\, , \,\, A_1(q^2)=F_0(q^2) \, .
\end{eqnarray}

In the following, we briefly outline  the operator product expansion for the correlation function $\Pi_{\mu\nu}(p_1,p_2)$  in perturbative
QCD.  We contract the quark fields in the correlation function
$\Pi_{\mu\nu}(p_1,p_2)$ with Wick theorem firstly,
\begin{eqnarray}
\Pi_{\mu\nu}(p_1,p_2)&=&\int d^4 x d^4 y e^{ip_2\cdot x-ip_1\cdot y} {\rm Tr} \left\{i\gamma_5C^{mn}(x)\gamma_{\mu}(1-\gamma_5)B^{nk}(-y)\gamma_\nu C^{km}(y-x) \right\}\, ,
\end{eqnarray}
replace the $c$ and $b$ quark propagators $C^{ij}(x) $ and $B^{ij}(x)$ with the corresponding full propagators $S_{ij}(x)$,
\begin{eqnarray}
S_{ij}(x)&=&\frac{i}{(2\pi)^4}\int d^4k e^{-ik \cdot x} \left\{
\frac{\delta_{ij}}{\!\not\!{k}-m_Q}
-\frac{g_sG^n_{\alpha\beta}t^n_{ij}}{4}\frac{\sigma^{\alpha\beta}(\!\not\!{k}+m_Q)+(\!\not\!{k}+m_Q)
\sigma^{\alpha\beta}}{(k^2-m_Q^2)^2}+\frac{\delta_{ij}\langle g^2_sGG\rangle }{12}\right.\nonumber\\
&&\left. \frac{m_Qk^2+m_Q^2\!\not\!{k}}{(k^2-m_Q^2)^4}+\cdots\right\} \, ,
\end{eqnarray}
 where $Q=c,b$, $t^n=\frac{\lambda^n}{2}$, the $\lambda^n$ are the Gell-Mann matrixes, the $i$, $j$ are color indexes, and the $\langle g^2_sGG\rangle$
is the gluon condensate \cite{Reinders85},
then carry out  the  integrals  with the help of the Cutkosky's rule. In this article, we take into account the leading-order perturbative contribution and  gluon condensate contributions in the operator product expansion, and show them explicitly using the Feynman diagrams   in Figs.1-2.

\begin{figure}
 \centering
 \includegraphics[totalheight=5cm,width=7cm]{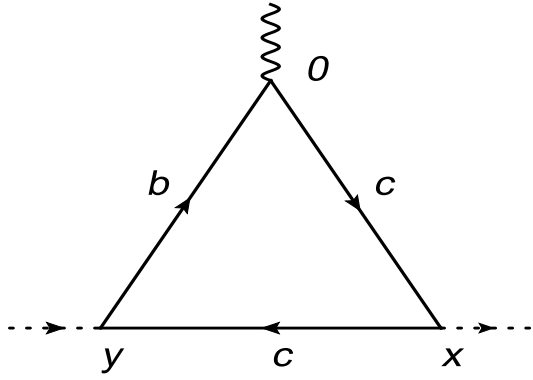}
    \caption{The leading-order perturbative contribution. }
\end{figure}

\begin{figure}
 \centering
 \includegraphics[totalheight=8cm,width=14cm]{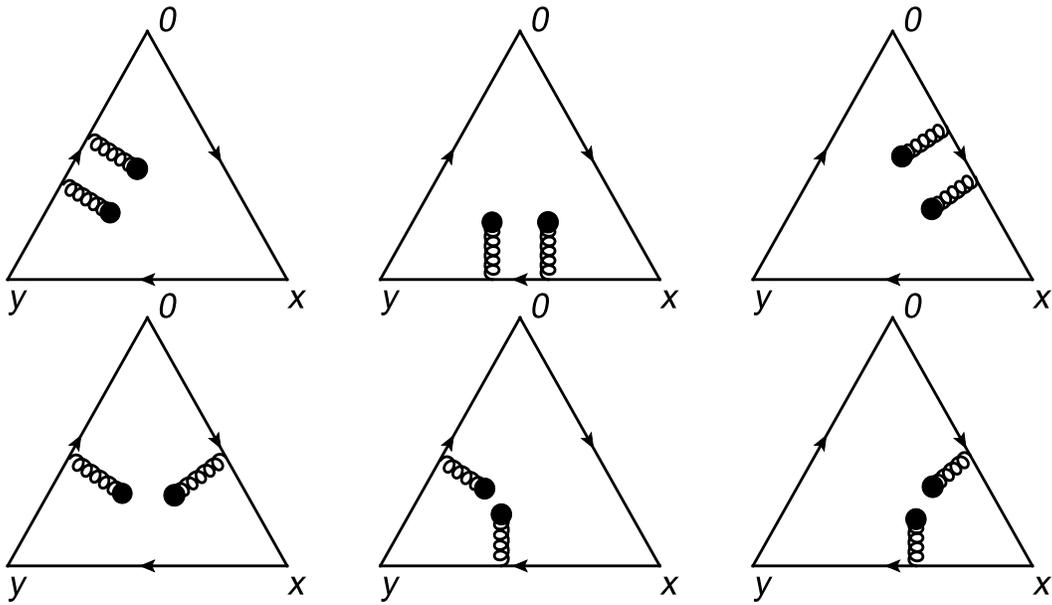}
    \caption{The gluon condensate contributions. }
\end{figure}

The leading-order contribution shown in Fig.1 can be written as
\begin{eqnarray}
\Pi_{\mu\nu}(p_1,p_2)&=&\frac{3}{(2\pi)^4}\int d^4k \frac{ {\rm Tr}\left\{ \gamma_5\left[ \!\not\!{k}+ \!\not\!{p}_2+m_c\right]\gamma_\mu(1-\gamma_5)\left[ \!\not\!{k}+ \!\not\!{p}_1+m_b\right]\gamma_\nu\left[ \!\not\!{k} +m_c\right]\right\}}{\left[(k+p_2)^2-m_c^2\right]\left[(k+p_1)^2-m_b^2\right]\left[k^2-m_c^2\right]}\, ,\nonumber\\
&=&\int ds_1 ds_2 \frac{\rho_{\mu\nu}(s_1,s_2,q^2)}{(s_1-p_1^2)(s_2-p_2^2)} \, .
\end{eqnarray}
We take the  following replacements to put all the quark lines on mass-shell using the Cutkosky's rule,
\begin{eqnarray}
\frac{ \!\not\!{k}+m_c}{k^2-m_c^2} &\to& -2\pi i \delta\left(k^2-m_c^2\right)(\!\not\!{k}+m_c) \, ,\nonumber\\
\frac{ \!\not\!{k}+ \!\not\!{p}_2+m_c}{(k+p_2)^2-m_c^2} &\to& -2\pi i \delta\left((k+p_2)^2-m_c^2\right)(\!\not\!{k}+ \!\not\!{p}_2+m_c) \, ,\nonumber\\
\frac{ \!\not\!{k}+ \!\not\!{p}_1+m_b}{(k+p_1)^2-m_b^2} &\to& -2\pi i \delta\left((k+p_1)^2-m_b^2\right)(\!\not\!{k}+ \!\not\!{p}_1+m_b) \, ,
\end{eqnarray}
and  obtain the leading-order perturbative spectral density  $\rho_{\mu\nu} (s_1,s_2,q^2)$,
\begin{eqnarray}
\rho_{\mu\nu}(s_1,s_2,q^2)&=& -\frac{3i}{(2\pi)^3} \int d^4k \delta\left[(k+p_2)^2-m_c^2\right]\delta\left[(k+p_1)^2-m_b^2\right]\delta\left[k^2-m_c^2\right]\nonumber\\
&& {\rm Tr}\left\{ \gamma_5\left[( \!\not\!{k}+ \!\not\!{p}_2)+m_c\right]\gamma_\mu(1-\gamma_5)\left[( \!\not\!{k}+ \!\not\!{p}_1)+m_b\right]\gamma_\nu\left[ \!\not\!{k} +m_c\right]\right\}\, .
\end{eqnarray}
 We calculate the Feynman diagrams shown in Fig.2 analogously with   the Cutkosky's rule,
the calculations are straightforward and tedious. In the following, we present the basic formulae  used in this article,
\begin{eqnarray}
\int d^4k\delta^3 &=&\frac{\pi}{2\sqrt{\lambda(s_1,s_2,q^2)}}\, ,  \\
\int d^4k k_\mu \delta^3&=&\frac{\pi}{2\sqrt{\lambda(s_1,s_2,q^2)}}\left[ a_1 p_{1\mu}+b_1 p_{2\mu}\right] \, , \\
\int d^4k k_\mu k_\nu\delta^3 &=& \frac{\pi}{2\sqrt{\lambda(s_1,s_2,q^2)}}\left[a_2p_{1\mu}p_{1\nu}+b_2p_{2\mu}p_{2\nu}+c_2(p_{1\mu}p_{2\nu}+p_{1\nu}p_{2\mu})+d_2g_{\mu\nu}\right]\, ,
\end{eqnarray}
where
\begin{eqnarray}
\delta^3 &=&\delta [k^2-m^2]\delta[(k+p_1)^2-m_1^2]\delta[(k+p_2)^2-m_2^2]\, ,\nonumber\\
a_1&=&-\frac{\widetilde{s}_2(s_1+s_2-q^2)-2s_2\widetilde{s}_1}{\lambda(s_1,s_2,q^2)} \, ,\nonumber\\
b_1&=&-\frac{\widetilde{s}_1(s_1+s_2-q^2)-2s_1\widetilde{s}_2}{\lambda(s_1,s_2,q^2)} \, ,\nonumber\\
a_2&=&\frac{\widetilde{s}_2^2+2s_2m^2}{\lambda(s_1,s_2,q^2)}+6s_2
\frac{s_1\widetilde{s}_2^2+s_2\widetilde{s}_1^2-\widetilde{s}_1\widetilde{s}_2(s_1+s_2-q^2)}{\lambda(s_1,s_2,q^2)^2}\, ,\nonumber\\
b_2&=&\frac{\widetilde{s}_1^2+2s_1m^2}{\lambda(s_1,s_2,q^2)}+6s_1
\frac{s_1\widetilde{s}_2^2+s_2\widetilde{s}_1^2-\widetilde{s}_1\widetilde{s}_2(s_1+s_2-q^2)}{\lambda(s_1,s_2,q^2)^2}\, ,\nonumber\\
c_2&=&\frac{1}{s_1+s_2-q^2}\left\{\frac{2\widetilde{s}_1\widetilde{s}_2(s_1+s_2-q^2)-3(s_1\widetilde{s}_2^2+s_2\widetilde{s}_1^2)}{\lambda(s_1,s_2,q^2)}
-m^2\left[1+\frac{4s_1s_2}{\lambda(s_1,s_2,q^2)}\right]\right.\nonumber\\
&&\left.-12s_1s_2
\frac{s_1\widetilde{s}_2^2+s_2\widetilde{s}_1^2-\widetilde{s}_1\widetilde{s}_2(s_1+s_2-q^2)}{\lambda(s_1,s_2,q^2)^2}\right\}\, ,\nonumber\\
d_2&=&\frac{m^2}{2}+\frac{s_1\widetilde{s}_2^2+s_2\widetilde{s}_1^2-\widetilde{s}_1\widetilde{s}_2(s_1+s_2-q^2)}{2\lambda(s_1,s_2,q^2)}\, ,
\end{eqnarray}
 $\widetilde{ s}_i=s_i+m^2-m_i^2$, $i=1,2$, and $\lambda(a,b,c)=a^2+b^2+c^2-2ab-2bc-2ca$. The formulae in Eqs.(20-21) are consistent with that obtained in Refs.\cite{Gongshi-Ioffe,Gongshi-Yang},
while the formula in Eq.(22) is slightly different from that of   Ref.\cite{Gongshi-Yang}.

 Once the analytical expressions of the correlation function at the quark level are obtained, then we can take  quark-hadron duality  below the threshold
$s^0_{1}$ and $s_2^0$ in the channels $B_c^*$ and $\eta_c$ respectively,
 take double Borel transform  with respect to the variables
$P_1^2=-p_1^2$ and $P_2^2=-p_2^2$ respectively,
finally   obtain four QCD  sum rules  for the weak form-factors,
\begin{eqnarray}
A_1(q^2)&=&\frac{2m_c}{f_{\eta_c}m_{\eta_c}^2f_{B_c^*}m_{B_c^*}(m_{B_c^*}+m_{\eta_c})} \int ds_1 ds_2 \left\{\frac{3\mathcal{C}\left[m_c(s_1+s_2-q^2)+s_2(m_b-m_c) \right]}{8\pi^2\sqrt{\lambda(s_1,s_2,q^2)}} \right. \nonumber\\
&&\left.+\frac{3m_c(s_1-q^2)+2m_bs_2}{12\pi\lambda(s_1,s_2,q^2)^{\frac{3}{2}}}\langle\frac{\alpha_sGG}{\pi}\rangle
  \right\} \exp\left\{\frac{m_{B_c^*}^2-s_1}{M_1^2}+\frac{m_{\eta_c}^2-s_2}{M_2^2}\right\}\, , \nonumber\\
\widetilde{A}_+(q^2)&=&\frac{2m_c(m_{B_c^*}+m_{\eta_c})}{f_{\eta_c}m_{\eta_c}^2f_{B_c^*}m_{B_c^*}} \int ds_1 ds_2 \left\{\frac{3\mathcal{C}}{4\pi^2\sqrt{\lambda(s_1,s_2,q^2)}} \right. \nonumber\\
&& \left[ m_c+(m_b-m_c)\frac{s_2^2-s_2(s_1+q^2+2m_c^2-2m_b^2)}{\lambda(s_1,s_2,q^2)}\right] \nonumber\\
&&+\frac{3s_1m_c}{2\pi\lambda(s_1,s_2,q^2)^{\frac{3}{2}}(s_1+s_2-q^2)}\left[1+\frac{4s_1s_2}{\lambda(s_1,s_2,q^2)}\right]\langle\frac{\alpha_sGG}{\pi}\rangle\nonumber\\
&&-\frac{m_c}{2\pi\lambda(s_1,s_2,q^2)^{\frac{3}{2}}(s_1+s_2-q^2)}\left[s_1+s_2+2q^2+\frac{12s_1s_2q^2}{\lambda(s_1,s_2,q^2)}\right]\langle\frac{\alpha_sGG}{\pi}\rangle\nonumber\\
&&+\frac{m_b}{3\pi\lambda(s_1,s_2,q^2)^{\frac{3}{2}}}\left[1+\frac{6s_1s_2}{\lambda(s_1,s_2,q^2)}\right]\langle\frac{\alpha_sGG}{\pi}\rangle -\frac{m_b}{3\pi\lambda(s_1,s_2,q^2)^{\frac{3}{2}}(s_1+s_2-q^2)}\nonumber\\
&&\left.\left[s_1-2s_2-q^2+\frac{6s_1s_2(s_1+s_2-q^2)}{\lambda(s_1,s_2,q^2)}\right]
\langle\frac{\alpha_sGG}{\pi}\rangle\right\}\exp\left\{\frac{m_{B_c^*}^2-s_1}{M_1^2}+\frac{m_{\eta_c}^2-s_2}{M_2^2}\right\}\, ,\nonumber\\
\widetilde{A}_-(q^2)&=&\frac{2m_c(m_{B_c^*}+m_{\eta_c})}{f_{\eta_c}m_{\eta_c}^2f_{B_c^*}m_{B_c^*}} \int ds_1 ds_2 \left\{\frac{3\mathcal{C}m_c\left[ 2s_1s_2-(s_1+m_c^2-m_b^2)(s_1+s_2-q^2)\right]}{2\pi^2\lambda(s_1,s_2,q^2)^{\frac{3}{2}}} \right. \nonumber\\
&&-\frac{3s_1^2m_c}{\pi\lambda(s_1,s_2,q^2)^{\frac{5}{2}} }\langle\frac{\alpha_sGG}{\pi}\rangle+\frac{m_c}{2\pi\lambda(s_1,s_2,q^2)^{\frac{3}{2}} }\left[1+\frac{6s_1q^2}{\lambda(s_1,s_2,q^2)}\right]\langle\frac{\alpha_sGG}{\pi}\rangle\nonumber\\
&&\left.-\frac{m_b}{3\pi\lambda(s_1,s_2,q^2)^{\frac{3}{2}}}\left[1+\frac{6s_1s_2}{\lambda(s_1,s_2,q^2)}\right]\langle\frac{\alpha_sGG}{\pi}\rangle \right\}\exp\left\{\frac{m_{B_c^*}^2-s_1}{M_1^2}+\frac{m_{\eta_c}^2-s_2}{M_2^2}\right\}\, ,\nonumber\\
V(q^2)&=&\frac{m_c(m_{B_c^*}+m_{\eta_c})}{f_{\eta_c}m_{\eta_c}^2f_{B_c^*}m_{B_c^*}} \int ds_1 ds_2  \left\{\frac{3\mathcal{C}}{4\pi^2\sqrt{\lambda(s_1,s_2,q^2)}} \right. \nonumber\\
&&\left. \left[ m_c+(m_b-m_c)\frac{s_2^2-s_2(s_1+q^2+2m_c^2-2m_b^2)}{\lambda(s_1,s_2,q^2)}\right] \right\}\exp\left\{\frac{m_{B_c^*}^2-s_1}{M_1^2}+\frac{m_{\eta_c}^2-s_2}{M_2^2}\right\} \, ,\nonumber\\
\end{eqnarray}
where
\begin{eqnarray}
\widetilde{A}_{+}(q^2)&=&A_{+}(q^2)+A_{-}(q^2)\, ,\nonumber\\
\widetilde{A}_{-}(q^2)&=&A_{+}(q^2)-A_{-}(q^2)\, ,\nonumber\\
\int ds_1 ds_2 &=&\int_{(m_b+m_c)^2}^{s_1^0}ds_1\int_{4m_c^2}^{s_2^0}ds_2\mid_{\mid 2s_1s_2-(s_1+s_2-q^2)(s_1+m_c^2-m_b^2) \mid \leq \sqrt{\lambda(s_1,s_2,q^2)\lambda(s_1,m_c^2,m_b^2)}}  \, , \nonumber\\
{\mathcal{C}}&=&\sqrt{\frac{4\pi\alpha_s^\mathcal{C}}{3v} \left[1-\exp\left(-\frac{4\pi\alpha_s^\mathcal{C}}{3v}\right)\right]^{-1}}\, ,\nonumber\\
v&=&\sqrt{1-\frac{4m_bm_c}{s_1-(m_b-m_c)^2}}\, .
\end{eqnarray}
For the heavy quarkonium state $B_c^*$, the relative velocity of quark movement is small, we should account for the Coulomb-like $\frac{\alpha_s^\mathcal{C}}{v}$  corrections.  After taking into account  all the Coulomb-like contributions shown in Fig.3, we obtain the coefficient $\mathcal{C}$ to dress the   quark-meson  vertex \cite{Coulomb-BC}. We take the approximation ${\alpha_s^\mathcal{C}}=\alpha_s(\mu)$ in numerical  calculations \cite{Wang-BC}.

\begin{figure}
 \centering
 \includegraphics[totalheight=5cm,width=6cm]{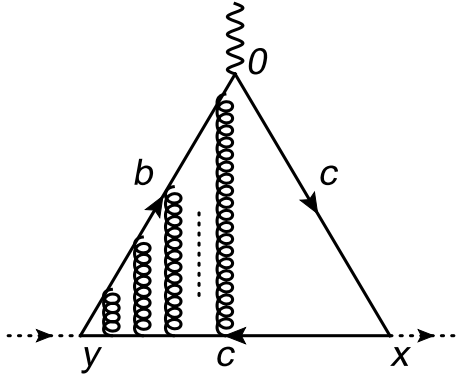}
    \caption{The ladder Feynman diagrams for the Coulomb-like interactions. }
\end{figure}

\section{Numerical results and discussions}
The hadronic input parameters are taken as $m_{\eta_c}=2.981\,\rm{GeV}$ \cite{PDG},  $s_1^0=(45\pm1)\,\rm{GeV}^2$, $f_{B_c^*}=0.384\,\rm{GeV}$, $m_{B^*_c}=6.337\,\rm{GeV}$ \cite{Wang-BC}, $s_2^0=(15\pm 1)\,\rm{GeV}^2$ \cite{SVZ79}, and $f_{\eta_c}=0.35\,\rm{GeV}$ \cite{PRT-1978}.
 The $B_c^{*\pm }$ mesons have not been observed yet, we take the  mass $m_{B^*_c}=6.337\,\rm{GeV}$ from the QCD sum rules \cite{Wang-BC}, which is consistent with
  the predictions of the relativized (or relativistic) quark models  \cite{GI,Raditive-width,GJ,ZVR},  nonrelativistic quark models \cite{Fulc,GKLT,EQ}, and  lattice QCD \cite{Latt}, see Table 1.
In the early work \cite{Khlopov-2},   Gershtein  and
 Khlopov obtained a simple relation $f_{ij}\propto m_i+m_j$ for the decay constant $f_{ij}$ of the pseudoscalar meson having the constituent quarks $i$ and $j$,
 the simple relation does not work well enough numerically. In this article, we take the values $f_{B_c^*}=0.384\,\rm{GeV}$ and $f_{\eta_c}=0.35\,\rm{GeV}$ from the QCD sum rules \cite{Wang-BC,PRT-1978}. The uncertainties of the weak  form-factors   originate from
the decay constants are $ \pm\frac{\delta f_{B_c^*}}{f_{B_c^*}} \pm \frac{\delta f_{\eta_c}}{f_{\eta_c}}$, therefore  the induced uncertainties of the radiative decay widths are  $ \pm2\frac{\delta f_{B_c^*}}{f_{B_c^*}} \pm 2\frac{\delta f_{\eta_c}}{f_{\eta_c}}$.
 The value of the gluon condensate $\langle \frac{\alpha_s
GG}{\pi}\rangle $ has been updated from time to time, and changes
greatly \cite{NarisonBook}, we use the recently updated value $\langle \frac{\alpha_s GG}{\pi}\rangle=(0.022 \pm
0.004)\,\rm{GeV}^4 $ \cite{gg-conden}.
For the heavy quark masses, we take the $\overline{MS}$ masses $m_{c}(m_c^2)=(1.275\pm0.025)\,\rm{GeV}$ and  $m_{b}(m_b^2)=(4.18\pm 0.03)\,\rm{GeV}$
 from the Particle Data Group \cite{PDG}, and take into account
the energy-scale dependence of  the $\overline{MS}$ masses from the renormalization group equation,
\begin{eqnarray}
m_c(\mu^2)&=&m_c(m_c^2)\left[\frac{\alpha_{s}(\mu)}{\alpha_{s}(m_c)}\right]^{\frac{12}{25}} \, ,\nonumber\\
m_b(\mu^2)&=&m_b(m_b^2)\left[\frac{\alpha_{s}(\mu)}{\alpha_{s}(m_b)}\right]^{\frac{12}{23}} \, ,\nonumber\\
\alpha_s(\mu)&=&\frac{1}{b_0t}\left[1-\frac{b_1}{b_0^2}\frac{\log t}{t} +\frac{b_1^2(\log^2{t}-\log{t}-1)+b_0b_2}{b_0^4t^2}\right]\, ,
\end{eqnarray}
  where $t=\log \frac{\mu^2}{\Lambda^2}$, $b_0=\frac{33-2n_f}{12\pi}$, $b_1=\frac{153-19n_f}{24\pi^2}$, $b_2=\frac{2857-\frac{5033}{9}n_f+\frac{325}{27}n_f^2}{128\pi^3}$,  $\Lambda=213\,\rm{MeV}$, $296\,\rm{MeV}$  and  $339\,\rm{MeV}$ for the flavors  $n_f=5$, $4$ and $3$, respectively  \cite{PDG}. In this article, we take the typical energy scale $\mu=2m_c(\mu^2)\approx 2\,\rm{GeV}$.

\begin{table}
\begin{center}
\begin{tabular}{|c|c|c|c|c|c|c|c|c|c|}\hline\hline
    \cite{GI} &\cite{Wang-BC} &\cite{Raditive-width}  &\cite{GJ} &\cite{ZVR} &\cite{Fulc} &\cite{GKLT} &\cite{EQ} &\cite{Latt}        \\ \hline
    6.338     &6.337          &6.332                  &6.308     &6.340      &6.341       &6.317       &6.337     &6.321         \\     \hline \hline
\end{tabular}
\end{center}
\caption{ The masses of the  $B_c^*$ mesons from different theoretical approaches, the unit is GeV. }
\end{table}

In Fig.4, we plot the weak form-factors at $q^2=0$ with variations of the Borel parameters $M_1^2$ and $M_2^2$ respectively. From the figure, we can see that the form-factors decrease monotonously with the increase of the Borel parameters  at the region $M_1^2\leq 4.0\,\rm{GeV}^2$ and $M^2_2\leq 2.5\,\rm{GeV}^2$, and no stable QCD sum rules can be obtained. In this article, we take the Borel parameters as $M_1^2=(5.0-7.0)\,\rm{GeV}^2$ and $M_2^2=(2.5-3.5)\,\rm{GeV}^2$, the values are rather stable with variations of the  Borel parameters. The contributions of high resonances and continuum states are greatly suppressed, $\exp(-\frac{s_1^0}{M_1^2})\leq e^{-6.3}$ and $\exp(-\frac{s_2^0}{M_2^2})\leq e^{-4.0}$. If we  choose much larger Borel parameters, the numerical values of the weak form-factors changes slightly, see Fig.4, the predictions still survive. The energy-scale $\mu^2$ and Borel parameters $M_1^2$, $M_2^2$ are of the same order, if we take the values $\mu^2=M_1^2=M_2^2=4\,\rm{GeV}^2$, the predictions change slightly.   The numerical values of the weak form-factors at zero momentum transfer are
 \begin{eqnarray}
A_1(0)&=&0.43\pm0.07 \, , \nonumber\\
A_+(0)&=&0.57\pm0.09 \, ,\nonumber\\
A_-(0)&=&0.85\pm0.15\, , \nonumber\\
V(0)&=&0.71\pm0.12\, .
\end{eqnarray}
If we take into account the uncertainty of the mass $m_{B^*_c}=6.337\pm 0.052\,\rm{GeV}$ from the QCD sum rules \cite{Wang-BC}, additional uncertainties
$\delta A_1(0)=\pm0.04$,  $\delta A_+(0)=\pm0.06$, $\delta A_-(0)=\pm0.09$, $\delta V(0)=\pm0.07$ are introduced, then
\begin{eqnarray}
A_1(0)&=&0.43\pm0.08 \, , \nonumber\\
A_+(0)&=&0.57\pm0.11 \, ,\nonumber\\
A_-(0)&=&0.85\pm0.17\, , \nonumber\\
V(0)&=&0.71\pm0.14\, .
\end{eqnarray}

From Eq.(24), we can also obtain the numerical values of the weak form-factors at the squared momentum $q^2$, then fit them  to an  exponential form,
  \begin{eqnarray}
  f(q^2)&=&f(0)\exp\left(c_1 q^2+c_2 q^4\right)\, ,
  \end{eqnarray}
   where the $f(q^2)$ denote the weak form-factors $A_1(q^2)$, $A_{+}(q^2)$, $A_{-}(q^2)$ and $V(q^2)$, the   $c_1$ are $c_2$ are fitted parameters.
 The numerical values of the fitted parameters $c_1$ and $c_2$ are presented in Table 2.

The calculations based on the three-point QCD sum rules indicate that  the $B_c \to \eta_c$ form-factor $F_1(0)$  is $0.20\pm0.02$ from Ref.\cite{Colangelo-BC}, $ 0.55\pm0.10$ from Ref.\cite{Narison-BC},
 $0.66$ from Ref.\cite{Coulomb-BC}, the discrepancies are rather large, as very different input parameters are taken in those studies. In the present work $A_1(0)\neq A_2(0)\neq V(0)\neq F(0)$, if the values of the $F(0)$ from Refs.\cite{Colangelo-BC,Narison-BC,Coulomb-BC} are taken, the heavy quark spin symmetry works not well enough, as the $c$ quark mass is not large enough.

\begin{figure}
 \centering
 \includegraphics[totalheight=6cm,width=7cm]{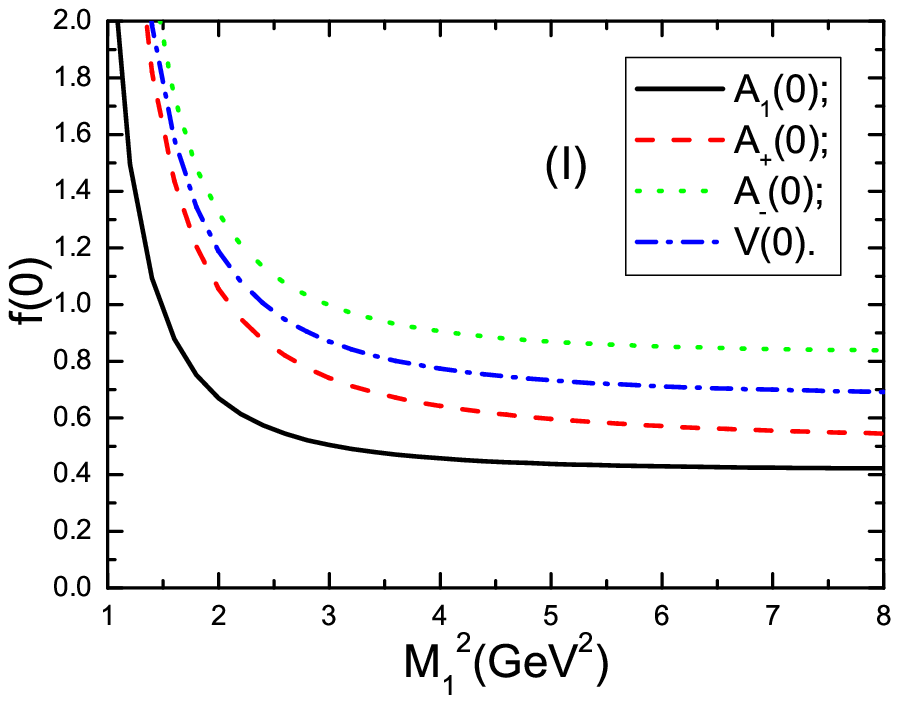}
\includegraphics[totalheight=6cm,width=7cm]{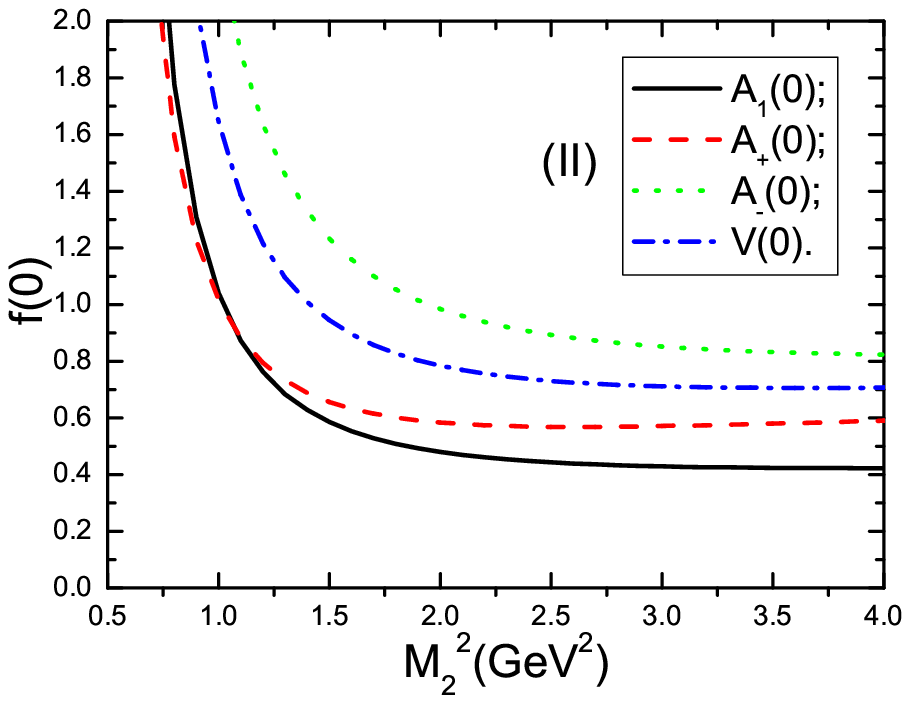}
        \caption{ The weak from-factors with variations of the Borel parameters $M_1^2$ and $M_2^2$, where
        $M_2^2=3.0\,\rm{GeV}^2$ in (I) and $M_1^2=6.0\,\rm{GeV}^2$ in (II). }
\end{figure}

The semileptonic decays $B_c^*\to \eta_c \ell \nu_{\ell}$ can be described by the effective Hamiltonian ${\cal H}_{\rm eff}$,
 \begin{eqnarray}
 {\cal H}_{\rm eff}&=&\frac{G_F}{\sqrt{2}}V_{cb} \bar{c}\gamma_{\alpha}(1-\gamma_5)b \,\bar{\nu}_{\ell}\gamma^{\alpha}(1-\gamma_5)\ell \, ,
 \end{eqnarray}
where the $V_{cb}$ is the CKM matrix element and the $G_F$ is the
Fermi constant.  We take into account the effective Hamiltonian  ${\cal H}_{\rm eff}$ and  the weak form-factors
$A_1(q^2)$, $A_{+}(q^2)$, $A_{-}(q^2)$ and $V(q^2)$
to obtain the squared amplitude $|T|^2$,
\begin{eqnarray}
|T|^2&=&4G_F^2 V_{cb}^2  (l^\alpha v^\beta+l^\beta v^\alpha-l\cdot v g^{\alpha\beta})\langle\eta_c(p)|j_\alpha(0) |B_c^*(P)\rangle \left[\langle\eta_c(p)|j_\beta(0) |B_c^*(P)\rangle\right]^{\dagger}\, ,
\end{eqnarray}
where the $P$, $p$, $l$ and $v$ are the four-momenta of the $B_c^*$, $\eta_c$, $\ell$ and $\bar{\nu}_{\ell}$, respectively. Finally
we obtain the differential  decay widths,
\begin{eqnarray}
d\Gamma&=&\sum\frac{|T|^2}{6m_{B_c^*}}\frac{dq^2}{2\pi}d\Phi(P\to q,p)\,d\Phi(q\to l,v) \, ,
\end{eqnarray}
where the $d\Phi(P\to q,p)$ and $d\Phi(q\to l,v)$  are the two-body phase factors defined  analogously, for example,
\begin{eqnarray}
d\Phi(P\to q,p)&=&(2\pi)^4\delta^4(P-q-p)\frac{d^3\vec{p}}{(2\pi)^3 2p_0}\frac{d^3\vec{q}}{(2\pi)^3 2q_0} \, .
\end{eqnarray}

We take  the revelent parameters as $G_F=1.166364\times 10^{-5}\,\rm{GeV}^{-2}$,  $V_{cb}=40.6\times 10^{-3}$, $m_{e}=0.510998928\,\rm{MeV}$,
 $m_{\mu}=105.6583715\,\rm{MeV}$, $m_{\tau}=1776.82\,\rm{MeV}$  from the Particle Data Group \cite{PDG},  then obtain the differential decay widths and
 decay widths,
\begin{eqnarray}
\Gamma(B_c^*\to \eta_c e \bar{\nu}_{e}) &=& 6.86^{+2.12}_{-1.83} {}^{+0.22}_{-0.21}{}^{+0.04}_{-0.04} {}^{+0.71}_{-0.65}\times 10^{-6}\,{\rm{eV}}\, , \nonumber \\
                                        &=& 6.86^{+2.25}_{-1.95} \times 10^{-6}\,{\rm{eV}}\, , \nonumber \\
\Gamma(B_c^*\to \eta_c \mu \bar{\nu}_{\mu}) &=& 6.84^{+2.11}_{-1.82}{}^{+0.22}_{-0.22}{}^{+0.04}_{-0.04} {}^{+0.71}_{-0.65}\times 10^{-6}\,{\rm{eV}}\, , \nonumber \\
                                            &=& 6.84^{+2.24}_{-1.95} \times 10^{-6}\,{\rm{eV}}\, , \nonumber \\
\Gamma(B_c^*\to \eta_c \tau \bar{\nu}_{\tau}) &=&2.15^{+0.66}_{-0.57}{}^{+0.05}_{-0.05}{}^{+0.04}_{-0.03}{}^{+0.35}_{-0.30}\times 10^{-6}\,{\rm{eV}}\, ,\nonumber \\
                                              &=&2.15^{+0.75}_{-0.65} \times 10^{-6}\,{\rm{eV}}\, ,
\end{eqnarray}
where the uncertainties  originate from the uncertainties of the $A_1(q^2)$, $A_{+}(q^2)$, $V(q^2)$ and $m_{B_c^*}$, sequentially.  The numerical values of the differential decay widths $d\Gamma/dq^2$ are shown in Fig.5. The  decay width of the radiative transition $B_c^* \to B_c\gamma$ is about tens of $\rm{eV}$ from the potential models \cite{Raditive-width}, the branching fractions of the $B_c^*\to \eta_c \ell \bar{\nu}_\ell$ are of the order $10^{-7}\sim 10^{-6}$. The tiny branching  fractions of the order $10^{-7}\sim 10^{-6}$ maybe   escape experimental  detections. The semileptonic decay widths of the $B_c$ mesons to charmonium states are also of the order $10^{-6}\,\rm{eV}$ \cite{Bc-semileptonic}, the corresponding branching fractions are of the order $10^{-3}$, as the $B_c$ mesons  have much smaller width $\Gamma_{B_c}=1.46\times 10^{-3}\,\rm{eV}$, the semileptonic decays of  the $B_c$ mesons to charmonium states are more easy to be observed. The $b\bar{b}$ pairs would be copiously produced at the LHCb \cite{LHC}, we expect that a large number of $B_c^*$ events would be accumulated, and the experimental study of the differential branching fractions of the semileptonic decays of $B^*_c$ mesons
 to charmonium states would be feasible. The differential branching fractions can be  measured as $ \Delta {\rm Br}/\Delta q^2$ in bins of the
momentum-transfer squared $q^2$. The LHCb collaboration has observed the first evidence for the hadronic annihilation decay  $B^{+} \to D_s^{+}\phi$  with significance more than $3\sigma$, the measured branching fraction  is ${\rm{Br}}(B^{+} \to D_s^{+}\phi) = \left(1.87^{+1.25}_{-0.73} \pm 0.19  \pm 0.32 \right) \times 10^{-6}$ \cite{LHCb-1210.1089}. The branching fractions ${\rm{Br}}(B^{+} \to D_s^{+}\phi)$ and ${\rm{Br}}(B_c^{*} \to \eta_c\ell\bar{\nu}_{\ell})$ are of the same order, we still expect that the $B_c^{*} \to \eta_c\ell \bar{\nu}_{\ell}$ be observed in the future at the LHCb.
 On the other hand, we can take
 the $B_c^* \to \eta_c$ form-factors as basic input parameters in the  phenomenological analysis of the two-body decays of the $B_c^*$ mesons, such as the
 $B_c^* \to \eta_c \pi$, $\eta_c \rho$, $\eta_c a_0(980)$, $\eta_c a_1(1260)$, $\eta_c a_2(1320)$, $\eta_c K$, $\eta_c K^*$, $\eta_c K_0(800)$, $\eta_c K_1(1270)$, $\eta_c K_1(1400)$, $\eta_c D$, $\eta_c D^*$, $\eta_c D_{0}$, $\eta_c D_{1}$, $\eta_c D_{2}$, $\eta_c D_s$, $\eta_c D^*_s$, $\eta_c D_{s0}$, $\eta_c D_{s1}$, $\eta_c D_{s2}$, etc.

\begin{table}
\begin{center}
\begin{tabular}{|c|c|c|c|c|}\hline\hline

   $A_1(0)$           & $A_{+}(0)$         & $A_{-}(0)$         & $V(0)$       \\ \hline
   $0.43\pm0.07$      & $0.57\pm0.09$      & $0.85\pm0.15$      & $0.71\pm0.12$        \\ \hline\hline
   $c_1/c_2$          & $c_1/c_2$          & $c_1/c_2$          & $c_1/c_2$      \\ \hline
   $0.0484/0.0000$    & $0.0710/0.0005$    & $0.0719/0.0004$    & $0.0715/0.0004$      \\ \hline\hline

\end{tabular}
\end{center}
\caption{ The parameters for the weak form-factors, the units of the $c_1$ and $c_2$ are $\rm{GeV}^{-2}$ and $\rm{GeV}^{-4}$, respectively. }
\end{table}

\begin{figure}
 \centering
 \includegraphics[totalheight=7cm,width=9cm]{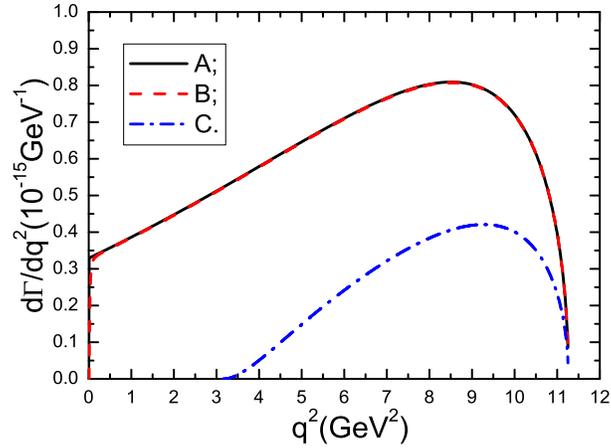}
    \caption{The differential decay widths with variations of the squared momentum $q^2$, the $A$, $B$ and $C$ denote
    $d\Gamma(B_c^*\to \eta_c e \bar{\nu}_{e})/dq^2$, $d\Gamma(B_c^*\to \eta_c \mu \bar{\nu}_{\mu})/dq^2$ and $d\Gamma(B_c^*\to \eta_c \tau \bar{\nu}_{\tau})/dq^2$, respectively. }
\end{figure}

\section{Conclusion}
In this article, we study the $B_c^* \to \eta_c$ form-factors with the three-point QCD sum rules, then take those weak form-factors as the basic input parameters
to calculate  the semileptonic decay widths and differential decay widths.   The tiny decay widths may be observed   experimentally  in the future at the LHCb,  while the $B_c^* \to \eta_c$ form-factors can be taken as basic input parameters in other  phenomenological analysis.

\section*{Acknowledgements}
This  work is supported by National Natural Science Foundation,
Grant Numbers 11075053, 11375063,  and the Fundamental Research Funds for the
Central Universities.

\end{document}